\documentclass[a4paper, amsfonts, amssymb, amsmath, reprint, showkeys, nofootinbib, twoside,floatfix]{revtex4-2}

\usepackage[english]{babel}
\usepackage[utf8]{inputenc}
\usepackage{array}
\usepackage{amsthm}
\usepackage{mathtools}
\usepackage{physics}
\usepackage{xcolor}
\usepackage{graphicx}
\usepackage[left=13mm,right=13mm,top=25mm,bottom=25mm,columnsep=15pt]{geometry} 
\usepackage{adjustbox}
\usepackage{placeins}
\usepackage[T1]{fontenc}
\usepackage{lipsum}
\usepackage{lipsum}
\usepackage{upgreek}
\usepackage{csquotes}

\usepackage{fancyhdr} 
\fancyheadoffset{0cm}

\fancypagestyle{plain}{%
  \fancyhf{}%
  \fancyhead[R]{\thepage}%
}

\usepackage{xr-hyper} 
\makeatletter
\usepackage[pdftex, pdftitle={Article}, pdfauthor={Author}]{hyperref}
\usepackage{epstopdf}
\usepackage{siunitx}
\usepackage{xcolor}
\usepackage{threeparttable}
\usepackage{tabularx}
\begin{document}
\raggedbottom

\title{High-efficiency graphene-silicon slot-waveguide microring modulator at 1.5\,\textmu m \textcolor{black}{and 2\,\textmu m wavelength bands}}
\author{Chao Luan$^{1,2*}$}
\author{Deming Kong$^{1}$}
\author{Yong Liu$^{1}$}
\author{Yunhong Ding$^{1*}$}
\author{Hao Hu$^{1*}$}

\affiliation{$^{1}$DTU Electro, Department of Photonics Engineering, Technical University of Denmark, DK-2800, Kgs. Lyngby, Denmark}

\affiliation{\textcolor{black}{$^{2}$Current address: Research Laboratory of Electronics, MIT, Cambridge, MA, 02139, USA}}

\affiliation{\textcolor{blue}{$*$chaoluan@mit.edu; yudin@dtu.dk; huhao@dtu.dk}}

\begin{abstract}

Electro-optic (E/O) modulators are crucial for optical communication but face a trade-off between modulation bandwidth and efficiency. A small footprint could reduce the capacitance and increase the bandwidth, however, this usually results in a low modulation efficiency. Here, we present an integrated E/O modulator that simultaneously achieves wideband large bandwidth and high modulation efficiency operation by embedding a partially overlapped double-layer graphene on a compact silicon slot waveguide microring resonator. \textcolor{black}{At 1550 nm, the graphene-silicon slot-waveguide demonstrates a high phase modulation efficiency of $V_{\pi}L$ = $220~\mathrm{V}\,\mu\mathrm{m}$, and the corresponding microring modulator has a large bandwidth of over 70 GHz, a compact active length of 10\,\textmu m, and an optical modulation amplitude (OMA) of -1.97 dBm under a 3-V voltage swing. The modulator operates at a data rate of 50 Gbit/s with an open eye diagram under a 2-V $V_\text{pp}$ RF drive voltage. The graphene modulator operation is broadband, and we also characterize its performance at 2\,\textmu m wavelength band. At 2\,\textmu m wavelength band, the microring modulator has a large bandwidth of over 20 GHz, an OMA of -3.36 dBm under a 6-V voltage swing, and an open eye diagram at 20 Gbit/s with a 2-V $V_\text{pp}$ RF drive voltage. The difference in performance is caused by the bandwidth limit of the 2\,\textmu m wavelength band measurement setup.} The \textcolor{black}{broadband,} large bandwidth, compact, highly efficient,  and energy efficient graphene E/O modulator has the potential to enable large-scale graphene photonic integrated circuits, facilitating a broad range of applications such as optical interconnects, optical neural networks, and programmable photonic circuits.

\end{abstract}
\maketitle
\section{Introduction}

The rapidly increasing demand for bandwidth for the Internet calls for fundamental breakthroughs in high-speed optical communication systems. Electro-optic (E/O) modulators perform an essential role in optical communication systems\cite{Reed2010,  Miller2017, Zhang2023} and crave for high modulation efficiency and large bandwidth. This demand is particularly significant for data centers, where there is a need for high speed, compact size, low cost, and low power consumption. Furthermore, emerging applications such as optical neural networks and programmable photonic circuits are anticipated to derive benefits from large bandwidth and highly efficient E/O modulators \cite{Bogaerts2020, shen2017, Dong2014, luan2026single, luan2025demonstration, luan2026large}. \textcolor{black}{To date, E/O modulators have been made in a number of platforms including silicon \cite{Zhang2023, yuan20245, Zhang2022, Sakib2021, timurdogan2014}, thin-film lithium niobate (TFLN)\cite{wang2018, He2019, Xue2022,Thomaschewski2020, Fang2023LiNbO3, Xu2020LiNbO3IQ, Renaud2023LiNbO3, Zhang2021LiNbO3Review}, plasmonic \cite{Eppenberger2023, Abel2019, Haffner2015, Haffner2018, Messner2019}, barium titanate (BaTiO$_3$)\cite{Abel2019}, Organic Electro-Optic polymer\cite{Palmer2014, Wolf2018, Benea2022OEOMetasurface, Moor2024PlasmonicTerabaud}, III-V material\cite{Han2017, Hiraki2017, Tang2012}, indium tin oxide (ITO)\cite{Amin2020}, and thin-film Lithium tantalate (TFLT)\cite{Wang2024LiTaO3, Nazar2025LiTaO3SiN, Zhang2025LiTaO3Comb, Wang2024LiTaO3Nature}.} Among these materials, \textcolor{black}{silicon shares the same material platform that enables complimentary-metal-oxide-semiconductor (CMOS) compatibility and high-volume manufacturing, TFLN modulator supports large bandwidth and low loss operation, plasmonic modulator delivers compact footprint and large bandwidth, polymer and III-V material-based E/O modulators are capable of operating at low driving voltages (below 1 V).} Despite these individual merits, simultaneously achieving large bandwidth and high modulation efficiency is challenging. Increasing the device footprint to accumulate stronger E/O interaction could lead to a higher modulation efficiency, but it would also result in a larger capacitance and consequently a smaller bandwidth. 

To address this challenge, we propose and demonstrate a microring modulator featuring a double-layer graphene on a silicon slot waveguide. The modulator is constructed with a partially overlapped double-layer graphene on a 50-nm-wide air slot waveguide that is embedded within a microring resonator. The graphene layers have a small length of only 10\,\textmu m and an overlapped width of only 110 nm. The silicon slot-waveguide structure provides both enhanced E/O interaction and reduced capacitance due to the small overlap between the top and bottom graphene layers, resulting in a high modulation efficiency and a large bandwidth simultaneously. The proposed graphene-silicon slot-waveguide demonstrates \textcolor{black}{strong phase modulation efficiency, with the $V_{\pi}L$ product of $220~\mathrm{V}\,\mu\mathrm{m}$ at 1.5\,\textmu m wavelength band, and the corresponding microring modulator exhibits a large bandwidth of over 70 GHz, an extinction ratio of 9.21 dB, an insertion loss of 1.41 dB, an OMA of -1.97 dBm under a 3 V voltage swing, and an open eye diagram at 50 Gbit/s with a 2-V $V_\text{pp}$ driving voltage.}

\begin{figure*}[t!]
\centering
\includegraphics[width=0.8\textwidth]{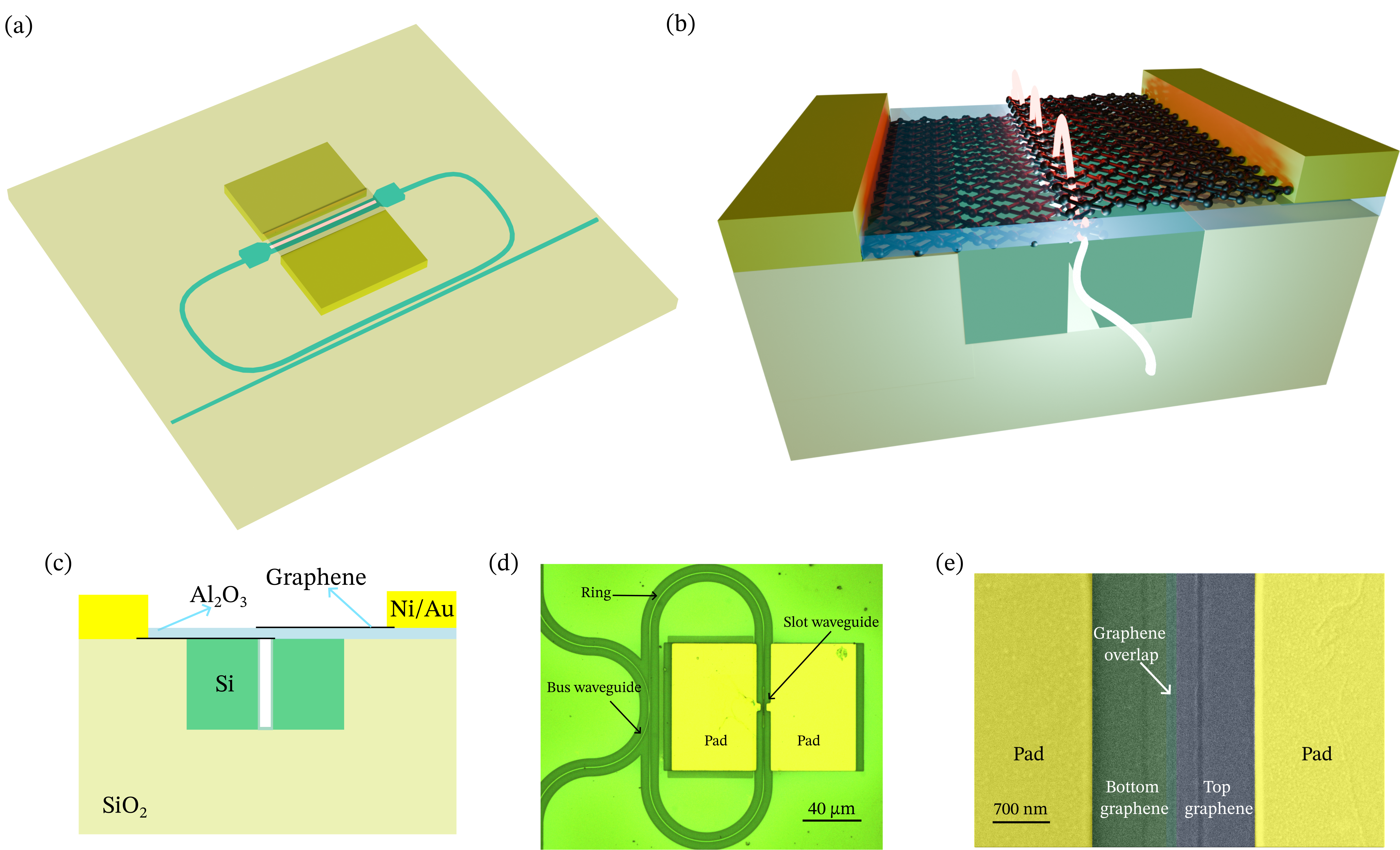}
\caption{Device concept. (a-c) Schematic and cross section of the graphene-silicon slot-waveguide microring E/O modulator. (d) Optical microscope image of the graphene-silicon slot waveguide microring modulator. (e) SEM image of the graphene-silicon slot waveguide microring modulator. }
\label{fig:1}
\end{figure*}

Our modulator employs graphene as the E/O material, which is promising due to its gapless band structure, high carrier mobility speed, and fast photocarrier generation/relaxation time. Additionally, graphene features low intrinsic optical loss, broadband operation, and CMOS compatibility \cite{Phare2015, Sorianello2018, Liu2011, Liu2012, Agarwal2021, Giambra2019, Mohsin2014, Dalir2016, Lee2021, Bonaccorso2010, Youngblood2014, Qiu2014, Hu2016, Datta2020, Xia2014, Ono2020, luan2022highpdp, luan2022integrated, luan2022high, luan2022ultra, luan20202}. However, conventional graphene E/O modulators typically suffer from weak light-graphene interaction and low modulation efficiency, and the relatively long length of the graphene limits the modulation bandwidth that can be achieved. To overcome these limitations, we propose a deep-subwavelength air-slot-waveguide microring structure. By placing the graphene on this ultra-narrow slot waveguide, the effective refractive index variation of the graphene is significantly enhanced. As a result, the length of the graphene is minimized, leading to a reduction in the effective device footprint and capacitance, which in turn enables a large E/O modulation bandwidth. 
A microring structure is used to recycle light, thereby further enhancing the modulation efficiency. However, combining both slot waveguide and microring structure is challenging due to the high loss caused by the slot waveguide, which typically results in a low quality factor and low efficiency of the microring modulator. Our device is designed to achieve a good balance between bandwidth and quality factor of the microring, thereby achieving high bandwidth and high modulation efficiency simultaneously. To reduce the loss \textcolor{black}{and increase the OMA}, the modulator works in the graphene transparency region, where the graphene features both strong electro-refraction and low optical loss. 

\textcolor{black}{Beyond the performance at the telecom C-band, we further demonstrate the broadband versatility of our graphene modulator by realizing graphene-based modulator operating at the mid-infrared wavelength of 2\,\textmu m band, which has become a promising candidate to be the next communication window (supplementary information VI). The 2\,\textmu m wavelength band modulator has a bandwidth of over 20 GHz (setup limited), an OMA of -3.36 dBm under a 6-V voltage swing, and delivers an open eye diagram at 20 Gbit/s with a 2-V $V_\text{pp}$ driving voltage (setup limited).}

We have also shown that our graphene modulator has high yield, which is crucial for developing large-scale photonic integrated circuits. We developed a new transfer process for graphene and optimized the device fabrication process (more details in Methods), enabling high-yield and high-quality transfer of large volumes of graphene. Our device fabrication process is compatible with CMOS technology and can be repeated with high consistency. These findings represent a significant advancement towards the realization of large-scale graphene-based photonic integrated circuits.

\section{Design}

\textcolor{black}{The microring resonator comprises three distinct structural sections. The first is the silicon strip waveguide, which serves as the bus-to-ring coupling region and the light propagation region in the ring. The second is the deep subwavelength silicon slot waveguide region where light modulation happens, and the third is the strip-to-slot waveguide mode converter where a 1×2 multimode interferometer (MMI) is employed. The MMI dimensions is 1.22\,\textmu m in width and 1.3\,\textmu m in length. Based on the self-imaging principle, the input field is reproduced as two images at the output of the MMI. A second, reversed MMI is placed at the opposite end of the slot section to convert the slot mode back to the strip mode. Fully etched photonic crystal grating couplers at 1.5\,\textmu m and 2\,\textmu m wavelength bands are designed to couple light into and out of the chip. The grating couplers are designed to have a coupling angle of 8$^{\circ}$ to match the commercial fiber array. Non-uniformed photonic crystal hole radius and pitch period are designed to maximize the coupling efficiency and achieve apodized coupling (supplementary information IV).}

\begin{figure*}[t!]
\centering
\includegraphics[width=0.65\textwidth]{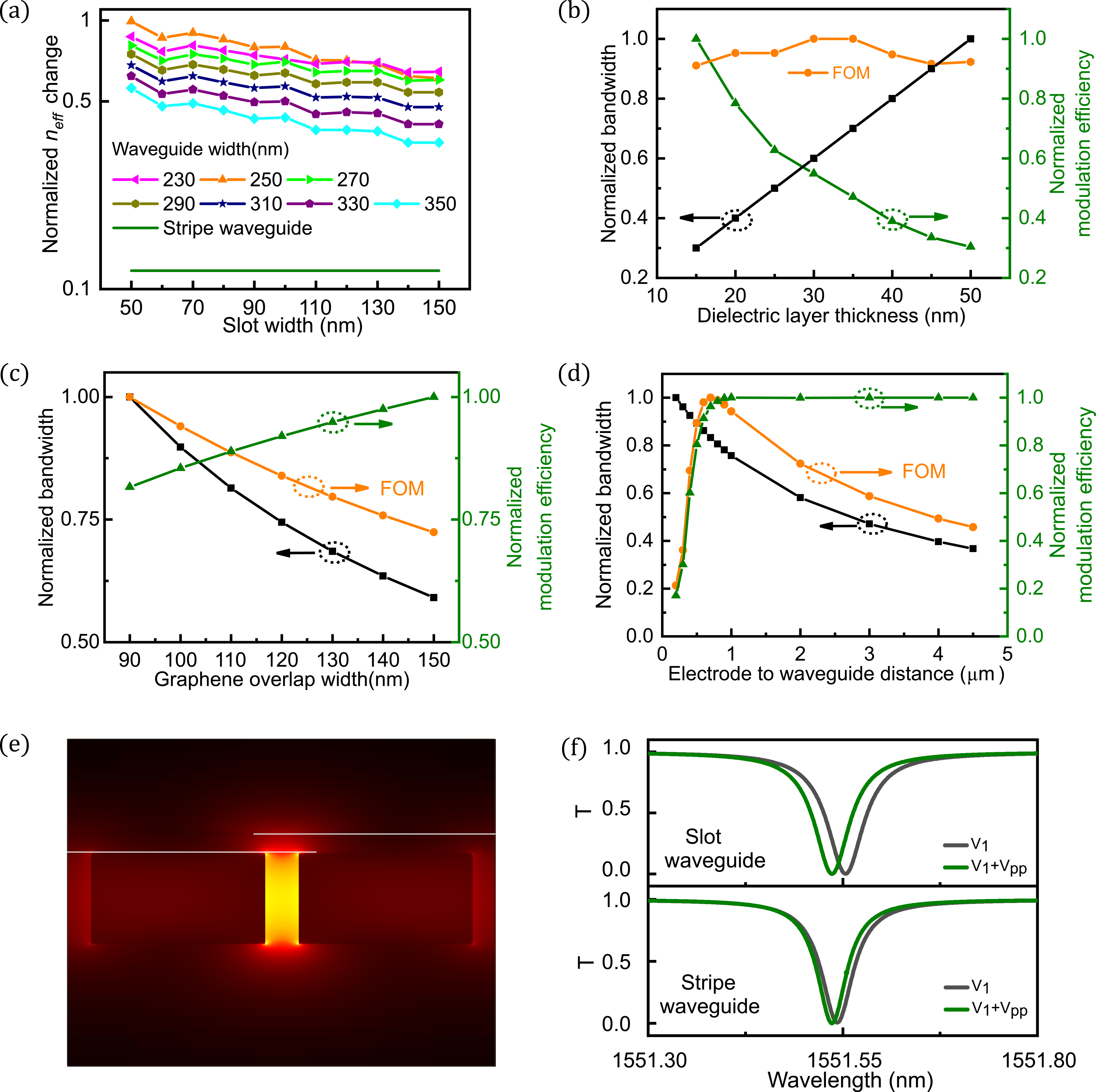}
\caption{Double-layer-graphene slot-waveguide E/O modulator optimization. (a) Calculated and normalized refractive index change $\Delta n_\text{eff}$ versus the slot waveguide dimension. (b) Calculated normalized modulation efficiency (green line), modulation bandwidth (black line) and modulation efficiency-bandwidth product (orange line) of the slot waveguide modulator versus dielectric layer thickness. The highest modulation efficiency-bandwidth product was obtained at the dielectric layer thickness of 35 nm. (c) Calculated normalized modulation efficiency (green line), modulation bandwidth (black line) and modulation efficiency-bandwidth product (orange line) of the slot waveguide modulator versus double layer graphene overlap width. (d) Calculated normalized modulation efficiency (green line), modulation bandwidth (black line) and modulation efficiency-bandwidth product (orange line) of the slot waveguide modulator versus electrode to waveguide distance. (e) Calculated eigenmode profile of the optimized slot waveguide. The light is tightly confined in the slot. (f) Calculated microring modulator transmission based on the slot and stripe waveguides.}
\label{fig:2}
\end{figure*}

Schematic of the deep subwavelength double layer graphene on silicon slot waveguide is shown in Fig.~\ref{fig:1} (a), (b) and (c). We use the finite-element method to calculate the optical modes and optimize the geometry of the slot waveguide. Light-graphene interaction is governed by the interband transition and intraband transition, which can be described by the complex surface conductivity model (supplementary information I). In the low Fermi level region, the interband transition plays the dominant role and the graphene mainly exhibits electro-absorption, while in the high Fermi level region, the intraband transition plays the dominant role and the graphene exhibits electro refraction (supplementary information I). Fig.~\ref{fig:2} (a) shows the normalized effective refractive index $\Delta n_\text{eff}$ under different slot waveguide dimensions, as indicated, the changes in the $\Delta n_\text{eff}$ are increased as the slot waveguide width decreases due to the field concentration effect. As a comparison, we also calculate the normalized refractive index change for the graphene-loaded silicon stripe waveguide. The slot waveguide exhibits more than an 8-fold increase of the \textcolor{black}{$\Delta n_\text{eff}$} compared with the stripe waveguide, which can be attributed to the enhanced mode overlapping and light-graphene interaction in the slot waveguide. A slot width of 50-nm and a waveguide width of 250-nm were designed, taking into account the fabrication processes.  

The tight light confinement and strong light-graphene interaction in the slot waveguide indicate that the device footprint required for light modulation can be greatly reduced, thereby making a small capacitance and a large bandwidth. Previously, the graphene modulator capacitance was reduced by using a thick dielectric layer, however, the large overlap width of the double-layer graphene and the large footprint result in a high device capacitance (several 10s fF). Here, the silicon slot waveguide design allows us to reduce the device capacitance by an order of magnitude through a partial graphene-overlap and compact device footprint. The key parameters for the modulator bandwidth are the thickness of the dielectric layer, the overlap width of the double-layer graphene, and the electrode to waveguide distance.

\begin{figure*}[t!]
\centering
\includegraphics[width=0.65\textwidth]{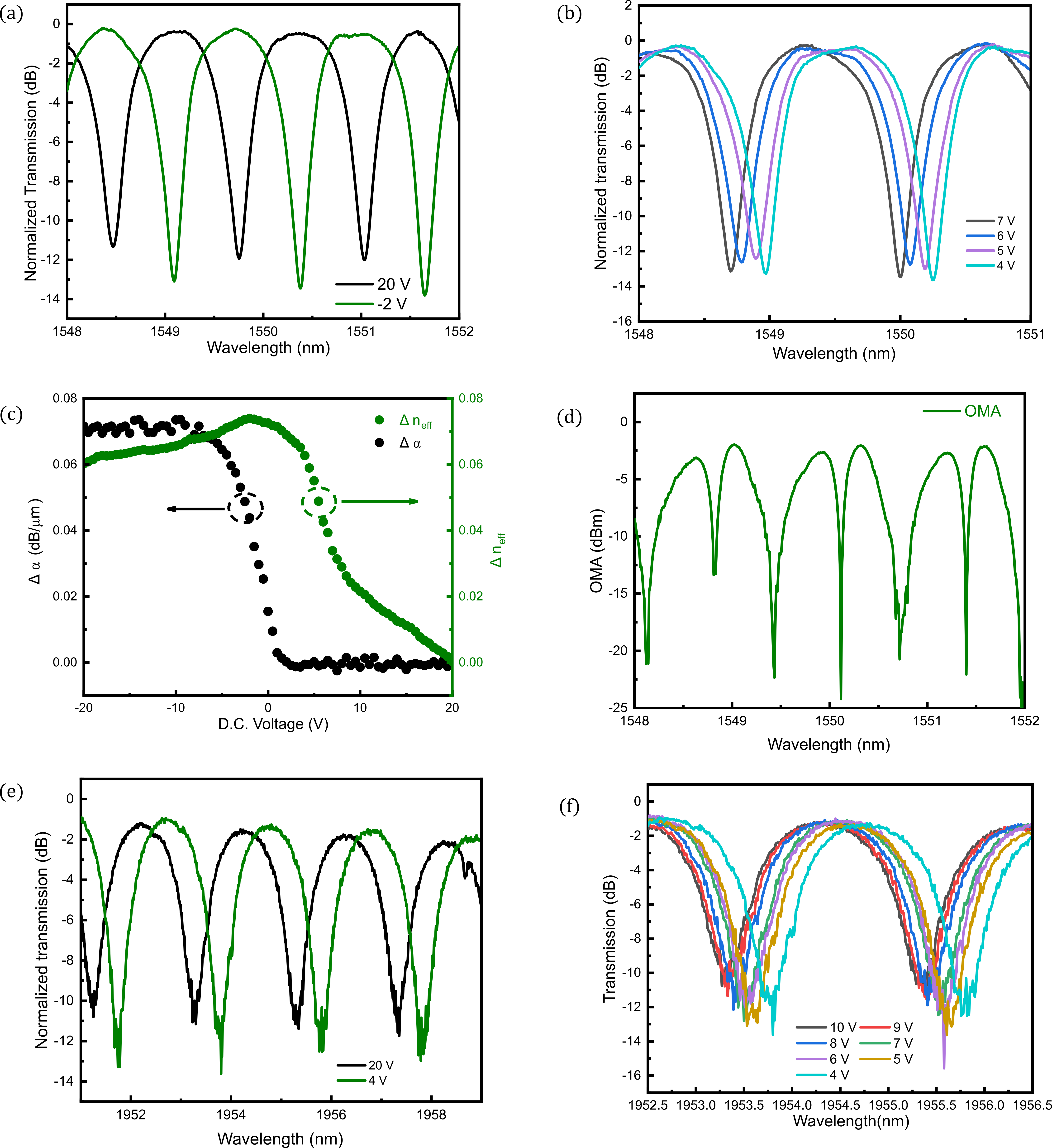}
\caption{Static Characterization of the double-layer-graphene slot-waveguide micro-ring modulator at 1.5\,\textmu m and 2\,\textmu m wavelength bands. (a) Optical transmission of the 1.5\,\textmu m waveband graphene microring modulator with 130-nm gap and 10\,\textmu m coupling length between the ring and bus waveguide, with the applied d.c. voltages of 20 V and -2 V, ${\pi}$ phase shift was obtained with 22-V voltage swing. (b) Optical transmission of the 1.5\,\textmu m waveband modulator at 4 V, 5 V, 6 V, and 7 V. (c) Calculated absorption coefficient and refractive index change of the 1.5\,\textmu m waveband graphene silicon slot waveguide modulator. (d) Calculated optical modulation amplitude of the graphene silicon slot waveguide modulator with a 3-V voltage swing at 5.5-V bias voltage. (e). Optical transmission of the graphene microring modulator at 2\,\textmu m wavelength band, 0.46${\pi}$ phase shift was obtained, with the applied d.c. voltages of 20 V and 4 V. (f). Optical transmission of the 2\,\textmu m waveband modulator at 5 V, 6 V, 7 V, 8 V, 9 V, and 10 V.}
\label{fig:3}
\end{figure*}

In addition to the modulation bandwidth, the modulation efficiency, is an important figure of merit for the E/O modulator, where the modulator needs to have large $\Delta n_\text{eff}$ in a compact footprint with low drive voltage and low insertion loss. Fig.~\ref{fig:2} (b) shows the normalized modulation efficiency (\textcolor{black}{defined as $\Delta n_\text{eff} / (\text{IL} \cdot L \cdot V_\text{pp})$, where $L$ is the active device length, $V_\text{pp}$ is the peak-to-peak drive voltage, and IL denotes the insertion loss}), modulation bandwidth and modulation efficiency-bandwidth product of the graphene-silicon E/O modulators as a function of the thickness of the dielectric layer. The modulation bandwidth increases while the modulation efficiency decreases as the thickness of the dielectric layer increases. The highest modulation efficiency-bandwidth product was achieved at a thickness of 35 nm for the dielectric layer, which provides a good balance between the bandwidth and efficiency. Fig.~\ref{fig:2} (c) shows the normalised modulation efficiency, the 3-dB bandwidth and the modulation efficiency-bandwidth product as a function of the overlap width of the double-layer graphene. When the overlap width was increased from 90 nm to 150 nm, the 3-dB bandwidth decreased 1.7-fold, however, the modulation efficiency only increased 1.2-fold. This can be attributed to the field concentration effect around the slot region \cite{Haffner2015, almeida2004guiding}. As a result, a narrow overlap is preferred to achieve a higher modulation efficiency-bandwidth product. Taking into account the fabrication tolerance and reproducibility of the fabrication processes, a double-layer graphene overlap width of 110 nm was employed (supplementary information II). 

Fig.~\ref{fig:2} (d) presents the calculated modulation efficiency, 3 dB modulation bandwidth and modulation efficiency-bandwidth product of the modulator as a function of electrode to waveguide distance for the designed silicon graphene slot waveguide modulator with a slot width of 50 nm, a dielectric layer thickness of 35 nm and a double layer graphene overlap width of 110 nm. There is a large decrease in the modulation efficiency when the electrode to waveguide distance is smaller than the evanescent decay length due to the increased loss. An electrode to waveguide distance of 700 nm was used, which provides the highest modulation efficiency-bandwidth product.

Based on these design, Fig.~\ref{fig:2} (e) shows the calculated electric field of the mode. As indicated, the electric field is strongly confined in the air slot (with 45\% of total power). We also calculated the transmission spectrum for the microring modulators based on the stripe waveguide and the slot waveguide. Fig.~\ref{fig:2} (f) shows the calculated transmission spectrum for a microring resonator with a graphene length of 10\,\textmu m. Compared to the stripe-waveguide based microring modulator, the slot-waveguide microring modulator exhibits an 8-fold increase in the \textcolor{black}{$\Delta n_\text{eff}$} and a 48-fold increase in the modulation efficiency (8-fold increase in the \textcolor{black}{$\Delta n_\text{eff}$} and 6-fold decrease in the insertion loss). 

\textcolor{black}{In addition to the optical waveguide geometry, the electrical design of the modulator was systematically optimized. The electrode follows a lumped-element approach, and the electrical design was optimized through the following strategies: (i) thick oxide underneath the pads—the silicon beneath the pads is etched away, and a high-resistivity SOI platform is used to reduce the pad-to-substrate parasitic capacitance; (ii) minimized routing metal length—the pads are connected directly to the graphene contact points, eliminating additional parasitic capacitance from long interconnect traces; (iii) small graphene–metal contact overlap of 700 nm to minimize parasitic capacitance at the contact interface; (iv) close electrode-to-waveguide placement, enabled by the strong optical confinement within the slot, which reduces the graphene sheet resistance in the access path and thereby lowers the total series resistance; and (v) rapid thermal annealing (RTA) to reduce the graphene–metal contact resistance. These design choices collectively minimize the RC time constant and enable the large electro-optic bandwidth of the modulator.}

\section*{Results and discussion}

The modulator was fabricated on a 220 nm thick high-resistivity silicon-on-insulator (SOI) platform (see methods). \textcolor{black}{Advanced silicon etching (ASE) was employed to realize the slot waveguide with a narrow width of 50 nm, a large aspect ratio of 4.4, and a smooth sidewalls that is critical for minimizing the scattering loss (supplementary information IV). The contact resistivity was reduced by using RTA, which led to a reduction of approximately 68\% in the contact resistance (supplementary information IV). A newly developed wet transfer process was employed to transfer a $2\,\text{cm} \times 2\,\text{cm}$ single-layer graphene sheet onto the chip (supplementary information IV).}

\begin{figure*}[t!]
\centering
\includegraphics[width=0.65\textwidth]{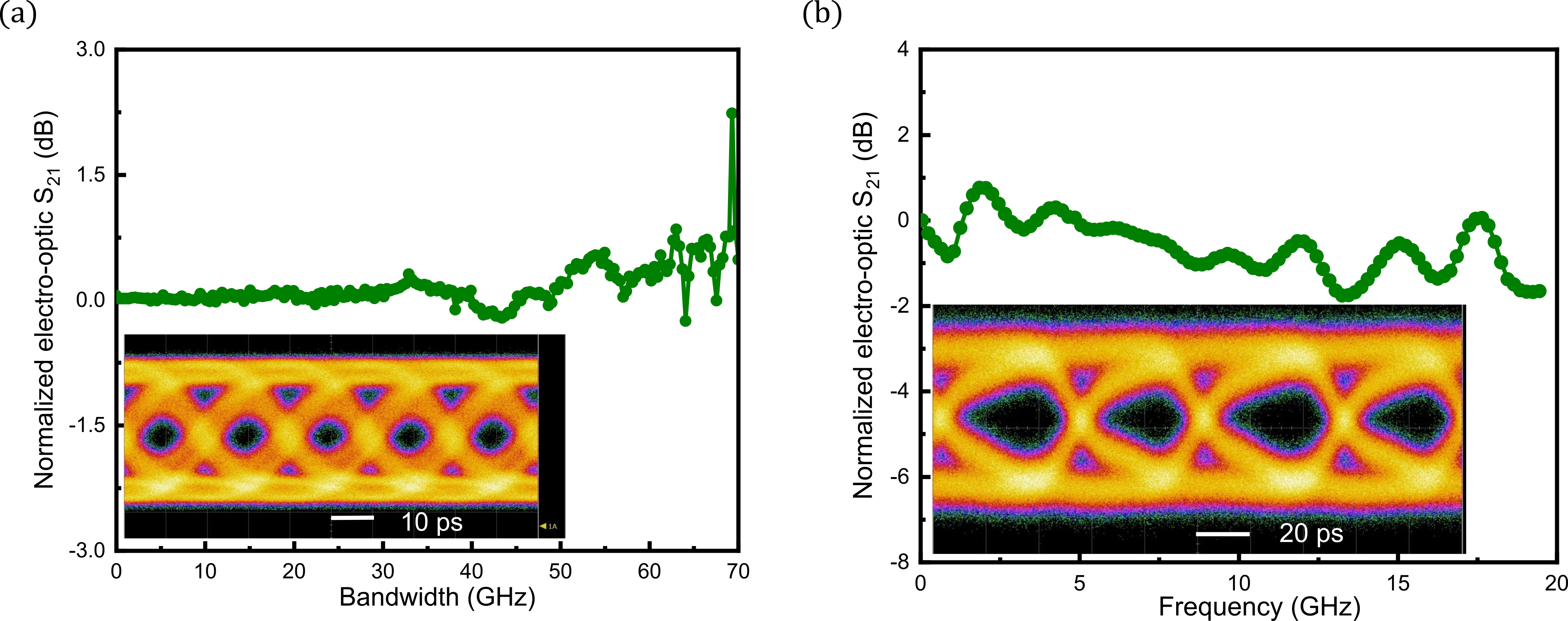}
\caption{High-speed characterization of the double-layer-graphene slot-waveguide microring modulator at 1.5\,\textmu m and 2\,\textmu m wavelength bands. (a) Measured electro-optic S$_{21}$ frequency response and the eye diagram of the modulator at 1.5\,\textmu m wavelength bands. The bandwidth of the modulator is beyond 70 GHz. Inset: measured 50 Gbit/s eye-diagram. (b) Measured electro-optic S$_{21}$ frequency response and the eye diagram of the modulator at 2\,\textmu m wavelength bands. The bandwidth of the modulator is beyond 20 GHz. Inset: measured 20 Gbit/s eye-diagram.}
\label{fig:4}
\end{figure*}

Fig.~\ref{fig:1} (d) shows the optical microscope image of the modulator, including the racetrack microring resonator and the buried bus waveguide. The ring resonator has a radius of 40\,\textmu m and a bus-ring coupling gap of 130 nm with a straight coupling length of 5\,\textmu m. Fig.~\ref{fig:1} (e) shows the artificial-colored scanning electron microscopy (SEM) image of the device, indicating the slot waveguide, the top and bottom layers of graphene, and the partial overlap of the double-layer graphene. \textcolor{black}{The 1.5\,\textmu m waveband and 2\,\textmu m waveband modulators have the same optical microscope and SEM images except for the grating couplers.}

\begin{figure*}[t!]
\centering
\includegraphics[width=0.35\textwidth]{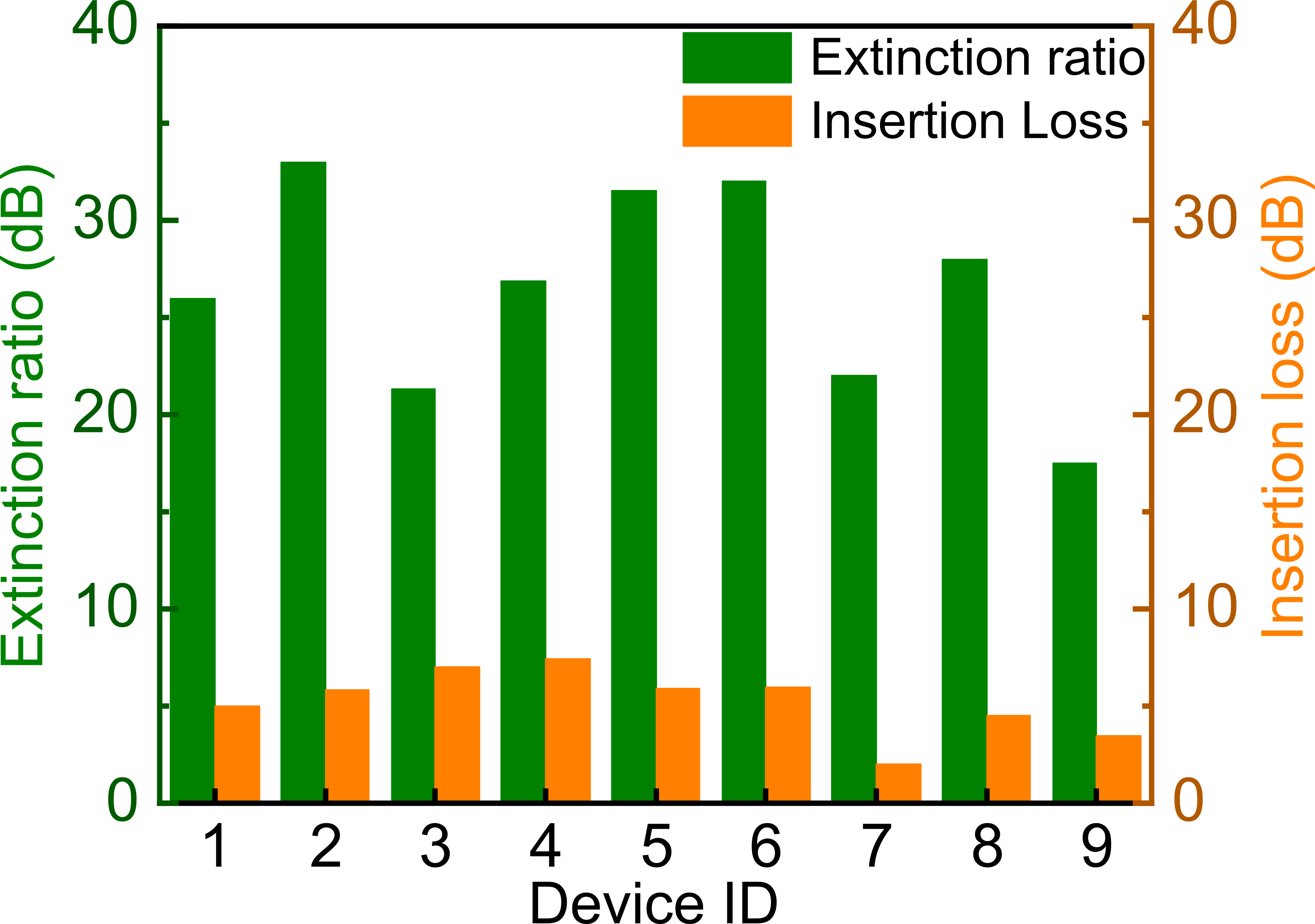}
\caption{Statistic of the performance of 10 graphene-silicon slot-waveguide modulators in terms of extinction ratio and insertion loss. The measurements are achieved from 10 devices with the slot width of 50 nm, bus-ring coupling length of 5\,\textmu m, bus-ring coupling gap ranging from 100 nm to 150 nm, and graphene length of 10\,\textmu m, respectively.}
\label{fig:5}
\end{figure*}

We measure the transmission spectrum of the E/O modulator at different bias voltages for \textcolor{black}{both 1.5\,\textmu m and 2\,\textmu m wavelength bands}. \textcolor{black}{The grating couplers were first characterized, the grating couplers have broad working bandwidth, and has a low loss of 4 dB at 1.5\,\textmu m, and 6.3 dB at 2\,\textmu m wavelength bands (supplementary information V). The MMI was specifically designed to work at 1.5\,\textmu m waveband, it has 0.2 dB loss at 1.5\,\textmu m waveband, and 0.5 dB loss at 2\,\textmu m waveband.} We sweep the DC voltage from -20 V to 20 V, and the microring modulators exhibit large resonance shifts in wavelength with a small change in amplitude at both wavelength bands (supplementary information V). \textcolor{black}{At 1.5\,\textmu m waveband, the graphene-silicon slot-waveguide achieves a full ${\pi}$ phase change with 10\,\textmu m length graphene under a 22-V voltage swing (Fig.~\ref{fig:3}) (a), while for the 2\,\textmu m wavelength band, the phase change is 0.46 ${\pi}$ (Fig.~\ref{fig:3} (e)). The corresponding $\Delta n_\text{eff}$ and $\Delta\alpha$ of 1.5\,\textmu m waveband modulator are calculated and plotted in Fig.~\ref{fig:3} (c), the $\Delta n_\text{eff}$ and $\Delta\alpha$ are nonlinear, and the modulator is working in the graphene transparency region. From the calculated $\Delta n_\text{eff}$ and $\Delta\alpha$ curve, we find the graphene is p doped (supplementary information V).} \textcolor{black}{Fig.~\ref{fig:3} (d) presents the modulator OMA as a function of wavelength with a 3-V voltage swing at a bias voltage of 5.5 V at 1.5\,\textmu m waveband. The modulators exhibit the largest OMA of -1.97 dBm. The 2\,\textmu m waveband measurement was performed using a thulium-doped fibre amplifier (TDFA) as the optical source,  which exhibits limited output power and reduced optical signal-to-noise ratio (OSNR). The OMA is -3.36 dBm at the 2\,\textmu m waveband, and is expected to be improved upon adoption of a high-power, low-noise 2\,\textmu m waveband tunable laser source. We didn't calculate the $\Delta n_\text{eff}$ and $\Delta\alpha$ at 2\,\textmu m waveband due to the limited resolution of the 2\,\textmu m waveband optical spectrum analyzer and the low OSNR of the transmission spectrum.}

The modulation bandwidth was measured using a vector network analyser (\textcolor{black}{Keysight N5227B VNA, 70-GHz bandwidth}) and a photodetector (\textcolor{black} {Finisar XPDV3120R, 70-GHz bandwidth for the 1.5\,\textmu m waveband measurement, and discovery DSC2-50S, 15-GHz bandwidth for the 2\,\textmu m waveband measurement}). At 1.5\,\textmu m waveband, the input laser wavelength was set to \textcolor{black}{1549.023 nm, which is the best OMA wavelength and is 45 pm detuning from the resonance wavelength.} A RF signal from the VNA was combined with a bias voltage and applied between the two graphene layers. The measured 3-dB bandwidth is over \textcolor{black}{70 GHz for the 1.5\,\textmu m modulator, which is larger than the photon lifetime (${\sim}$60\,GHz for $Q = 3226$) due to the wavelength detuning of the ring resonator}. \textcolor{black}{At 2\,\textmu m waveband, a CW laser with center wavelength of 2000 nm was employed and the measured bandwidth is over 20 GHz, which is limited by the available bandwidth of the photodetector. The calculated resistor-capacitor circuit limited bandwidth is over 100 GHz, thanks to the very small graphene capacitance of ${\sim}$2.5 fF (supplementary information IV). }

\textcolor{black}{The modulator can be driven with a CMOS compatible RF driving voltage. A 50\,Gbit/s} electrical signal was generated by a bit pattern generator with a pseudo-random binary sequence (PRBS) of $2^{15}-1$ length and amplified to \textcolor{black}{a $V_\text{pp}$ of 2\,V} with a 5\,V DC bias. As shown in Fig.~\ref{fig:4} (a), the open eye diagram of \textcolor{black}{50\,Gbit/s} at 1.5\,\textmu m waveband indicates a high-quality modulated signal, and the power consumption ($CV^2/4$) of the graphene microring modulator is ${\sim}$10.5\,fJ/bit. \textcolor{black}{At 2\,\textmu m waveband, open eye diagram of 20\,Gbit/s was obtained, which is limited by the photodetector bandwidth.}

The high-modulation-efficiency graphene modulator also features high reproducibility. Fig.~\ref{fig:5} shows the reproducibility of the graphene-silicon slot-waveguide microring modulators. These modulators feature an average extinction ratio of 26.4 dB, an average insertion loss of 5.2 dB, \textcolor{black}{an average OMA of -5.3 dBm under a 6-V voltage swing}, and the bandwidths are all above 40 GHz at resonance wavelengths. These statistics were obtained from measurements taken from 10 devices. High reproducibility is extremely important for future mass production and for the applications of large-scale photonic integrated circuits such as optical interconnects, programmable photonic circuits, and optical neural networks.

To summarize, we have demonstrate a broadband E/O modulator with large bandwidth and high modulation efficiency \textcolor{black}{at both 1.5\,\textmu m and 2\,\textmu m wavelength bands.} Our modulator features a 10\,\textmu m length double-layer graphene on a 50-nm wide silicon slot waveguide. \textcolor{black}{At 1.5\,\textmu m waveband, the modulator achieves a bandwidth over 70 GHz, an OMA of -1.97 dBm, and delivers an open eye diagram at 50 Gbit/s with 10.5 fJ/bit energy consumption. At 2\,\textmu m waveband, the modulator has a bandwidth over 20 GHz, an OMA of -3.36 dBm, and an open eye diagram at 20 Gbit/s was demonstrated, which is a proof of principle demonstration and paves a route toward an efficient transceiver in the 2\,\textmu m window.} The high reproducibility of the graphene modulators represents an important milestone towards practical graphene devices. Moreover, our integrated graphene E/O modulator platform could inspire a new generation of ultrafast large-scale integrated optoelectronic circuits. Our proposed concept of a double-layer graphene on a slot waveguide is also compatible with other platforms such as SiN and TiO$_\text{2}$ and has potential for other applications including photodetectors and sensors. Overall, our results represent a significant advancement towards the development of high performance graphene-based photonic integrated circuits.

\section*{Methods}

\textcolor{black}{The graphene microring modulator was fabricated on a commercial high resistivity silicon-on-insulator (SOI) chip with a 220-nm-thick silicon layer on top of a 2-\textmu m-thick SiO$_2$ buried layer. E-beam lithography and inductively coupled plasma etching are used to fabricate the 50-nm-wide slot waveguides and other passive components. To ensure a high graphene transfer-quality yield, PECVD SiO$_2$ was deposited and planarized to the top surface of the waveguide by using standard chemical mechanical polishing (CMP) technique to provide a flat surface which otherwise tends to break the graphene across the waveguide edge while drying. A spacer of 5\textasciitilde7-nm-thick Al$_2$O$_3$ was then uniformly atomic layer deposited on the surface of the waveguide to isolate the carrier transportation between graphene and silicon waveguide. CVD graphene on copper foil was spin-coated with 1~\textmu m AZ5214E resist, baked at \SI{90}{\degreeCelsius} for 2~minutes until dry, then soaked in a homemade copper etching solution (hydrochloric acid:DI water = 1:7, with a few drops of hydrogen peroxide) for 24~h and rinsed thoroughly in deionized (DI) water. To remove metal residuals, a RCA clean step-two solution was employed before the wet-transfer process. The transferred graphene was left to dry for one week. The graphene pattern was then defined using e-beam lithography and O$_2$ plasma etching. Next, the metal pads were defined using UV lithography and deposited by e-beam evaporation following a lift-off process with 10~nm Ni and 100~nm Au. A rapid thermal annealing (RTA) process was then used to significantly improve graphene--pad conductivity and reduce contact resistance. During the RTA process, the sample was ramped to \SI{450}{\degreeCelsius} in \SI{30}{\second} and held at \SI{450}{\degreeCelsius} for approximately \SI{1}{\minute} under a flowing gas mixture of 10\% hydrogen in nitrogen, with the cycle repeated five times. Direct deposition of high-dielectric-constant materials on pristine graphene by ALD is challenging due to the hydrophobic nature of the graphene basal plane. A 1~nm Al seed layer was thermally evaporated and quickly oxidized into Al$_2$O$_3$ upon exposure to air, after which 35~nm Al$_2$O$_3$ was deposited at \SI{200}{\degreeCelsius} by ALD. A similar process was performed for the top graphene layer and top metal pad as for the bottom layer. The top Al$_2$O$_3$ layer above the grating and metal pad was opened via H$_3$PO$_4$ wet etching.}

\section*{Data Availability}
The data from this work is stored in the server computer (DTU-CZC8028T24) at DTU and will be made available upon reasonable request.

\renewcommand{\bibsection}{\section*{REFERENCES}}
\bibliography{main}

\section*{Acknowledgements}
The author thanks Emily Theobald from MIT for helping polishing the schematic drawing, and Dr. Peixiong Shi, Dr Thomas Pedersen from DTU Nanolab for the equipment training. 

\subsection*{Author Contribution}
C.~Luan,  Y.~Ding and H.~Hu conceived the project. C.~Luan led the overall effort, including device design, simulation, fabrication, and characterization. C.~Luan also prepared the original manuscript and led the revision process, including additional data acquisition, analysis, and preparation of the revised manuscript. D.~Kong assisted with the high speed eye-diagram measurements. Y.~Liu assisted with the device fabrication including the slot waveguide etching recipe developing. C.~Luan wrote the manuscript with the feedback from  all the authors. Y.~Ding and H.~Hu co-supervised the project. This work was supported by a research grant (15401) of Young Investigator Program (2MAC) and QUANPIC (00025298) from VILLUM FONDEN.

\section*{Competing Interests}
The author declares no competing interests.

\end{document}